**Annular Modes of Variability in the Atmospheres of Mars and Titan**

J. Michael Battalio* and Juan M. Lora

Department of Earth and Planetary Sciences

Yale University

210 Whitney Ave.

New Haven, CT 06511

*Corresponding Author:  michael@battalio.com

Annular modes explain much of the internal variability of Earth's atmosphere but have never been identified on other planets. Using reanalyses for Mars and a simulation for Titan, we demonstrate that annular modes are prominent in the atmospheres of both worlds, explaining large fractions of their respective variabilities. One mode describes latitudinal shifts of the jet on Mars, as on Earth, and vertical shifts of the jet on Titan. Another describes pulses of midlatitude eddy kinetic energy on all three worlds, albeit with somewhat different characteristics. We further demonstrate that this latter mode has predictive power for regional dust activity on Mars, revealing its usefulness for understanding Martian weather. In addition, our finding of similar annular variability in dynamically diverse worlds indicates its ubiquity across planetary atmospheres, opening a new avenue for comparative planetology as well as an additional consideration for characterization of extrasolar atmospheres.

Annular modes arise from the internal dynamics of the atmosphere[1,2]. In Earth's atmosphere, they explain much of the weekly to monthly variability of the jet stream, synoptic wave activity, and precipitation[3–10] so are vital to understanding and predicting weather patterns. They are linked to the position and strength of the jet stream and atmospheric storm track. Modes are "annular" if the variability they represent is zonally symmetric—that is, if changes occur in tandem along whole latitude circles[11]. Two types of annular modes exist within the atmosphere of Earth. The first mode arises as north-south vacillations (quantified as the first empirical orthogonal function [EOF]; see Methods) in the anomalous surface pressure[1,3] or zonal-mean zonal wind[4,6]. This mode appears independently in both the northern and southern hemispheres, in each case explaining 20–30% of the variance in the zonal-mean wind, geopotential height, and surface pressure[1]. The mode physically represents shifts in atmospheric mass between the polar regions and the middle latitudes. (The second EOF of the zonal-mean wind, representing the next most variance, describes strengthening of the jet in place[6,12].) The vertical uniformity of the wind signature and importance of momentum fluxes associated with this mode indicate it is a barotropic feature[4,6,8,13,14]. The second type of annular mode—the Baroclinic Annular Mode—dominates the variance of zonal-mean eddy kinetic energy (henceforth EKE) and also appears in both the northern and southern hemispheres, explaining 43% and 34%, respectively, of the EKE variance[8,14,15]. This mode associates with anomalous eddy heat fluxes and varies with height, indicating a baroclinic nature.

The importance of annular variability on Mars was heretofore thought to be minimal[16,17], despite Mars having similar atmospheric dynamics[18], atmospheric energy cycles[19], baroclinic wave life-cycles[20–22], and semiannual oscillation[23] as Earth. Conversely, Titan inhabits a different regime of atmospheric dynamics, with a high Rossby number[24] (though near-surface baroclinic waves occur in simulations of Titan's atmosphere[25–28]). Motivated by the possibility that annular modes may help diagnose and predict weather on other planets as they do for Earth, we investigate if annular modes—including searching for baroclinic modes for the first time—are present on Mars and Titan. This would indicate their ubiquity in terrestrial atmospheres and reveal an important common source of atmospheric variability for consideration on both Solar System and extrasolar planets.

The purportedly inconsequential influence of barotropic annular modes on Mars[16,17] belied issues with the horizontal weighting of the analyzed fields (see Methods). The seemingly minor change to the correct weighting has enormous impacts on the resulting importance of annular modes on Mars. Using EOF analysis with the correct weighting, we identify both baroclinic and barotropic annular modes on Mars and Titan and also compare them to Earth's modes. For Mars, we focus on each hemisphere's fall and winter seasons and analyze two reanalysis datasets, the Mars Analysis Correction Data Assimilation (MACDA)[29] and the Ensemble Mars Atmosphere Reanalysis System (EMARS)[30]. For Titan, we evaluate the full year using simulations from the Titan Atmosphere Model[31].

**Martian annular mode in zonal wind**

Indeed, our EOF analysis of the atmosphere of Mars (see Methods) demonstrates many annular features reminiscent to those on Earth. Much of the large-scale variability of the mid-to-high latitude atmospheric flow is explained by an annular mode in the zonal-mean zonal wind (henceforth U-AM) (Fig. 1a–d and Extended Data Fig. 2a–d). In addition, we identify for the first time an annular mode in the zonal-mean EKE (EKE-AM) on another planet (Fig. 2a–d). These modes are the equivalents of Earth's barotropic (Fig 3 of ref. 4, Fig. 2a of ref. 6, Fig. 2a of refs. 8, 14) and baroclinic (Fig. 2f of refs. 8, 14) annular modes, respectively.

Two spatial structures dominate the variability of Mars's zonal-mean zonal wind. A dipolar structure—which is equivalent to Earth's U-AM—straddles the region of strongest winds (annual-mean zonal wind contoured in Fig. 1a–d) and explains the most variance in the northern hemisphere (~30–40%, Fig. 1b, d). As on Earth, the dipolar pattern represents latitudinal shifts of the jet. Because topography in the southern hemisphere varies greatly with latitude, such north-south shifts are disrupted, and additional spatial structures are also important in that hemisphere: A mono-polar structure—with a single center of action—accounts for slightly more variance (~25–35%, Extended Data Fig. 2a, c) than the dipolar structure (~20–30%, Fig. 1a, c), indicating that both are similarly important. Regardless, the spatial locations of the dipolar modes (Fig. 1a–d) align in both hemispheres and across datasets, with positive centers at 70º N/S and negative centers at approximately 40º N/S, all at around 100 Pa (approximately 20 km altitude), indicating that the spatial pattern is robust.

The dipolar U-AM behaves like Earth's U-AM, which links to eddy momentum fluxes and lacks any tilt in the vertical—a barotropic structure. This inter-planetary similarity of the barotropic mode is revealed by regressing the eddy momentum fluxes onto the U-AM at a -1 day lag (Fig.

1i–l). Convergence of the momentum fluxes maintains changes in the zonal-mean wind. Poleward eddy momentum fluxes occur between the zonal-mean zonal wind dipole; thus, divergence of the momentum flux occurs in the easterlies (negative pole), and convergence of momentum fluxes occurs in the westerlies (positive pole), demonstrating the barotropic nature of the U-AM and closely resembling Earth's zonal-wind annular mode (cf. Figs. 2a of refs. 8, 14). Poleward eddy momentum fluxes emerge in all datasets, with the northern hemisphere MACDA producing the largest anomalous eddy momentum fluxes, again indicating this is a robust structure of the Martian atmosphere. (The equatorial fluxes exhibited by EMARS in the northern hemisphere above 100 Pa may be associated with the stronger retrograde jet in EMARS[32].) The U-AM regresses only weakly on the eddy fluxes of heat at a lag of -1 day (Fig. 1e–h), similarly to the case with Earth's mode in zonal-mean wind[8,14], which indicates the U-AM is not a baroclinic feature. This demonstrates that variability in the zonal-mean wind arises from similar dynamics on both Earth and Mars and proves that annular modes are important dynamical features of the Martian atmosphere.

Like Earth's barotropic mode, Mars's U-AM is truly annular. The former can be defined in either the surface pressure or the zonal-mean zonal wind (cf. Fig. 6 of ref. 10). Comparing the U-AM calculated using either the zonal-mean wind or the surface pressure yields similar results for Mars as well. Calculating the U-AM from the daily-mean surface pressure, with the correct weighting (see Methods), duplicates the annular structure (Extended Data Fig. 1a) found when first defining the U-AM from zonal-mean zonal wind (as was done for Fig. 1) and then regressing back onto the surface pressure (Extended Data Fig. 3c). Both structures equally reflect the movement of mass away from the pole at times of westerly anomalies in the jet. If this analysis is repeated with an incorrect horizontal weighting, the annular structure is relegated to EOF 3, as in previous work[16,17] (Extended Data Fig. 1f). Thus, as on Earth, the methods for defining the U-AM on Mars, using either the zonal-mean winds or the surface pressure, are equivalent.

**Martian annular mode in eddy kinetic energy**

Just as the U-AM resembles Earth's barotropic mode, the newly-identified Martian EKE-AM resembles Earth's baroclinic annular mode (cf. Figs. 2f of ref. 8, 14, and Fig. 1 of ref. 15). The first spatial pattern is a single monopole that overlaps the greatest EKE and explains between 48 and 65% of the EKE variance (Fig. 2a–d), far larger than for Earth[8,14]. The location of the regressed zonal-mean EKE matches that of Earth's baroclinic mode, but the magnitude of the EKE-AM is approximately double. The annular feature achieves a maximum at 60–70º N/S and 10–150 Pa and is robust across both hemispheres and both datasets. Given the large degree of importance Earth's mode has in explaining extratropical wave activity, the comparatively larger percentage of variance explained by the EKE-AM points to its immense influence in determining wave activity, and therefore dust events[33], on Mars.

The EKE-AM links to poleward eddy heat fluxes that vary vertically (Fig. 2e–h), which matches the behavior of Earth's baroclinic mode and demonstrates that the EKE-AM represents (baroclinic) instabilities that instigate the type of traveling waves that initiate large dust events[33]. The relationship between eddy heat fluxes and the EKE-AM holds across hemispheres and datasets excepting the southern hemisphere MACDA domain, wherein fluxes are weaker. Nevertheless, each domain exhibits a peak in the magnitude of eddy heat fluxes near the surface in the midlatitudes, with a secondary peak around 100 Pa slanted poleward and passing through the maximum of the EKE (Fig. 2e–h, contours).

Earth's baroclinic and barotropic annular modes are decoupled, meaning that there is essentially no correlation between them; they act independently of one another. Additionally, Earth's mode in EKE is not linked to eddy fluxes of momentum, and Earth's mode in zonal wind is not linked to eddy fluxes of heat[8,14]. The Martian U-AM follows this pattern, as it does not regress strongly on either eddy heat fluxes or EKE (Fig. 1e–h). However, this is not the case for the Martian EKE-AM: The EKE-AM is associated with eddy fluxes of momentum (Fig. 2i–l) that are approximately double those related to the U-AM (Fig. 1i–l), though the EKE-AM does not regress on zonal-mean zonal wind (Fig. 2i–l, contours). This is distinctly different from Earth's EKE-AM[8,14]; thus, the Martian EKE-AM cannot be established as strictly a baroclinic mode. The entangled nature of Martian annular modes corroborates wave analyses[22] that find that transient eddies grow barotropically as well as baroclinically, depending on the period of the dominant waves. Thus, while annular modes on Mars are quite similar to those of Earth, the unique conditions of Mars provide some intriguing differences that demand continued investigation.

The connections between the EKE-AM and dust storm-producing traveling waves on Mars are corroborated by comparing the storm tracks on Mars to the EKE-AM. The mass-integrated EKE[33], which is a measure of the intensity of waves, regressed on the EKE-AM peaks at 45–75º N and 15–60º S (Fig. 3c, d). It connects to the EKE-AM upstream of Acidalia, Arcadia, and

Utopia Planitiae in the northern hemisphere, and near Argyre and Hellas Basins in the southern hemisphere, with only minor disagreements in magnitude between datasets (Extended Data Fig. 3e–h). Indeed, each of these regions hosts areas of increased storm activity from transient waves[22,33,34,40]. Furthermore, the clear annular structure, with longitudinal localization in storm tracks, is remarkably reminiscent of Earth's EKE-AM, which also pinpoints the Pacific, Atlantic, and Southern Ocean storm tracks (Fig. 3a, b)[8,14,15]. The similarity in the annular modes in EKE on both Earth and Mars therefore suggests that the EKE-AM may be essential for midlatitude weather on Mars.

**Impact of Martian annular modes on dust activity**

Earth's annular modes link to observable and impactful atmospheric features, like precipitation[15], further implying that the newfound EKE-AM might be expected to impact observable Martian weather like dust activity. To probe the link between the northern hemisphere EKE-AM and dust activity, we regress the EKE-AM onto the Mars Dust Activity Database[35] for the dusty season of one Mars year. Regions where northern hemisphere dust storms initiate, like Acidalia, Utopia, or Arcadia Planitae, are highlighted when dust activity leads the EKE-AM in the regression (Fig. 4a), demonstrating a relationship between the mode and observable, impactful surface conditions on Mars. That the EKE-AM pinpoints regions related to dust storm activity in an independent dataset also increases the confidence that annular modes on Mars truly exist. In these three regions in the northern hemisphere, dust activity peaks before the EKE-AM, meaning that dust is lifted and begins to be transported by atmospheric waves before the waves reach their peak in-tensity. This is analogous to the relationship between Earth's EKE-AM and precipitation, where precipitation peaks one day before waves do[15].

A key difference between precipitation on Earth and dust on Mars is that the latter remains lofted for weeks after initially being lifted. Thus, with Mars, the impact of the EKE-AM remains long after the peak of the EKE itself. In fact, when the dust activity lags the EKE-AM (i.e., the EKE-AM peak precedes the dust), locations into which northern hemisphere dust storms evolve—or flush[33,35,36]—link positively to the northern EKE-AM (Fig. 4b). This relationship between the northern EKE-AM and southern hemisphere dust activity maximizes near 0.12 storms/sol between Argyre and Hellas Basins, which is the region into which most dust storms travel[33,35]. Therefore, the leading behavior of the EKE-AM for activity that moves from the north to the south indicates predictive abilities of annular modes for dust storms. Only those dust events that flush into the southern midlatitudes become large enough to impact the surrounding atmosphere[33], so the EKE-AM could indicate when large dust events are favored before they occur. These sorts of predictions could be vital, for example, to ensuring the safety of future crewed missions to Mars.

**Titan's annular modes**

We have shown that Earth and Mars share similar annular modes in zonal wind and EKE, which could be due to their similarity in dynamical regime. To test this, we also investigate annular modes on Titan, which presents a heretofore unexplored dynamical regime from the perspective of annular variability. With key differences from Earth and Mars, Titan, too, supports annular modes of variability in zonal-mean zonal wind and EKE, based on simulations with a Titan general circulation model[31] (no reanalysis exists yet for Titan; see Methods). While potentially inaugurating additional research into Titan's atmospheric variability, these analyses should be interpreted with caution since they are from a single, albeit well-validated[26,27,31], model. Nevertheless, evidence of annular modes on Titan, in addition to Earth and Mars, points to the ubiquity of annular variability across the Solar System, which may hold significant promise for understanding atmospheric behavior across worlds.

The U-AM in our Titan simulations is dipolar but with the opposing poles stacked vertically (Fig. 5a, b) instead of horizontally like on Earth or Mars. Titan's U-AM explains ~68% of the variance of the zonal-mean zonal wind, far more than for Earth or Mars. The negative pole resides at 300 hPa (approximately 30 km altitude), while the weaker positive pole resides near the lowest vertical extent of the jet: This spatial structure represents vertical oscillations of the jet, as opposed to the horizontal vacillations of the terrestrial and Martian U-AM. Titan's U-AM is characterized by co-located zonal wind and poleward momentum fluxes (Fig. 5i, j). There is little relationship between the U-AM and EKE, but the U-AM is associated with poleward heat fluxes near the surface and equatorward heat fluxes in the mid-levels (Fig. 5e, f). Despite the vertical alignment of the U-AM dipole, the mode is associated with low surface pressures at the pole and an annular feature between 30 and 45º N/S in the anomalous mass-integrated EKE (not shown), which may indicate that the waves associated with the EKE have barotropic components, in agreement with expectations for Titan's dynamical regime[37]. This is counter to the nature of the terrestrial and Martian U-AM.

Titan's EKE-AM, though similar to Mars's and Earth's modes, also provides intriguing differences. Titan's EKE-AM explains 38.5 and 52% of the southern and northern hemisphere variance, respectively, in the zonal-mean EKE and has a single center of action at 500 hPa and 60º N/S (Fig. 5c, d). However, this mode exhibits a vertically stacked dipole of eddy heat fluxes that are poleward at high altitudes and equatorward in the mid-levels. These equatorward eddy heat fluxes indicate that the waves generating the EKE cannot be baroclinic in the mid-levels of the atmosphere[38]. Additionally, the EKE-AM regresses only weakly on the zonal wind (Fig. 5k, l, contours), similarly to Earth's baroclinic mode, but links strongly to eddy momentum fluxes (Fig. 5k, l), like the Martian EKE-AM (Fig. 2). Titan's annular modes, therefore, share characteristics of both Earth's and Mars's modes. Whether they too have predictive power for weather on Titan remains to be explored.

**Implications and Perspectives**

Previous studies suggested that annular variability might be unimportant in the atmosphere of Mars[16,17]; however, we find that annular variability is even more important on Mars than it is on Earth. The northern and southern Martian modes explain larger percentages of variance in the zonal wind than Earth's U-AM (Fig. 1). And, for the first time, we identify a baroclinic annular mode in the EKE on Mars and on Titan as well, both of which explain almost double the amount of variance in EKE—and thus wave activity—compared to Earth's baroclinic mode.

That Mars and Titan both exemplify annular variability opens a new window for comparative planetology and climatology. Mars's modes are more similar to Earth's, but the possible modes of Titan demonstrate how the influence of differing planetary parameters modifies the atmospheric dynamics. Annular structure in the EKE-AM occurs on Earth, Mars, and Titan between 35 and 65º N/S as seen in mass-integrated EKE (Fig. 3). These six fields are strikingly similar given the disparate topographies, storm tracks, and—particularly for Titan—dynamical regimes of the three worlds. Mars's EKE-AM amplifies within the three northern[21,34,39] and two southern[22,40] storm tracks that generate dust activity (Fig. 3c,d). Titan's EKE-AM controls EKE throughout an annulus between 35 and 60º N/S, and activity is most strongly indicated in the northern hemisphere between 180 and 300º E (Fig. 3e, f), suggesting that there may be preferred locations for eddy activity as on Earth and Mars. In addition to longitudinal localization of the storm tracks for Earth and Mars, mass-integrated EKE is dramatically amplified throughout the mid-latitudes on all three worlds (Fig. 3).

An exciting application lies in the use of annular dynamics to explain observable phenomena on Mars, Titan, and elsewhere. Annular modes on Earth link to the distribution of precipitation[15], cloud cover[9], and trace atmospheric species[5,7]. For Mars, dust is the most important visible consequence of changing weather patterns, and the Martian EKE-AM indeed exhibits a strong link to dust activity (Fig. 4). Further, the high degree of annularity of the Martian polar vortex[32] may relate to the relatively more important Martian U-AM and EKE-AM. In addition, connecting the EKE-AM to dust activity, and thus atmospheric temperatures[41], provides a physical (as opposed to a dynamical[12]) coupling to the Martian U-AM via changes in the surface pressure by sublima-tion/deposition of the $CO_2$ seasonal ice cap. For Titan, annular modes may play a role in the sporadic nature convective events[42], but further observations and improved model simulations will be required to assess their relationship.

Finally, the existence of annular modes in atmospheres other than Earth's proves the ubiquity of annular modes of variability across planetary atmospheres and demands a search for these modes beyond Mars and Titan. For example, Venus exhibits annularity in the "cold collar" of temperatures surrounding the pole[43]; this may link to annular variability as Earth's annular modes link to low temperatures[8]. Given this ubiquity, annular variability might be expected in gas and ice giants as well, for instance given the likely importance of eddies in driving Jupiter's jets[44]. Annular modes might also impart intrinsic variability that sets the noise floor for detections of exoplanet winds via Doppler shifts[45] so should be understood in that context. Each of these possibilities opens a promising avenue of research toward understanding how annular dynamics controls observable variability in the atmospheres of Earth and other planets.

**Figure 1. Spatial signature of the dipolar annular mode in anomalous zonal-mean zonal wind on Mars for both reanalysis datasets. (a–d)** Time-averaged zonal-mean zonal wind (contours every 10 m/s) and regressions of the dipolar mode onto anomalous zonal-mean zonal wind (shading). Both pairs of columns show the southern (left) and northern (right) hemispheres for the reanalysis indicated. The individual column titles indicate which EOF corresponds to the dipolar mode and give the percent of variance explained. Topography is shaded in gray. **(e–h)** Regressions of the mode onto anomalous zonal-mean eddy kinetic energy (contours every 10 m$^2$/s$^2$) and the anomalous eddy heat flux at -1 sol lag (shading). **(i–l)** Contours duplicated from shading in the top row (contours every 0.5 m/s, with dashed contours indicating negative values) and the anomalous eddy momentum flux at -1 sol lag (shading). Only regressions exceeding 99% confidence are shown.

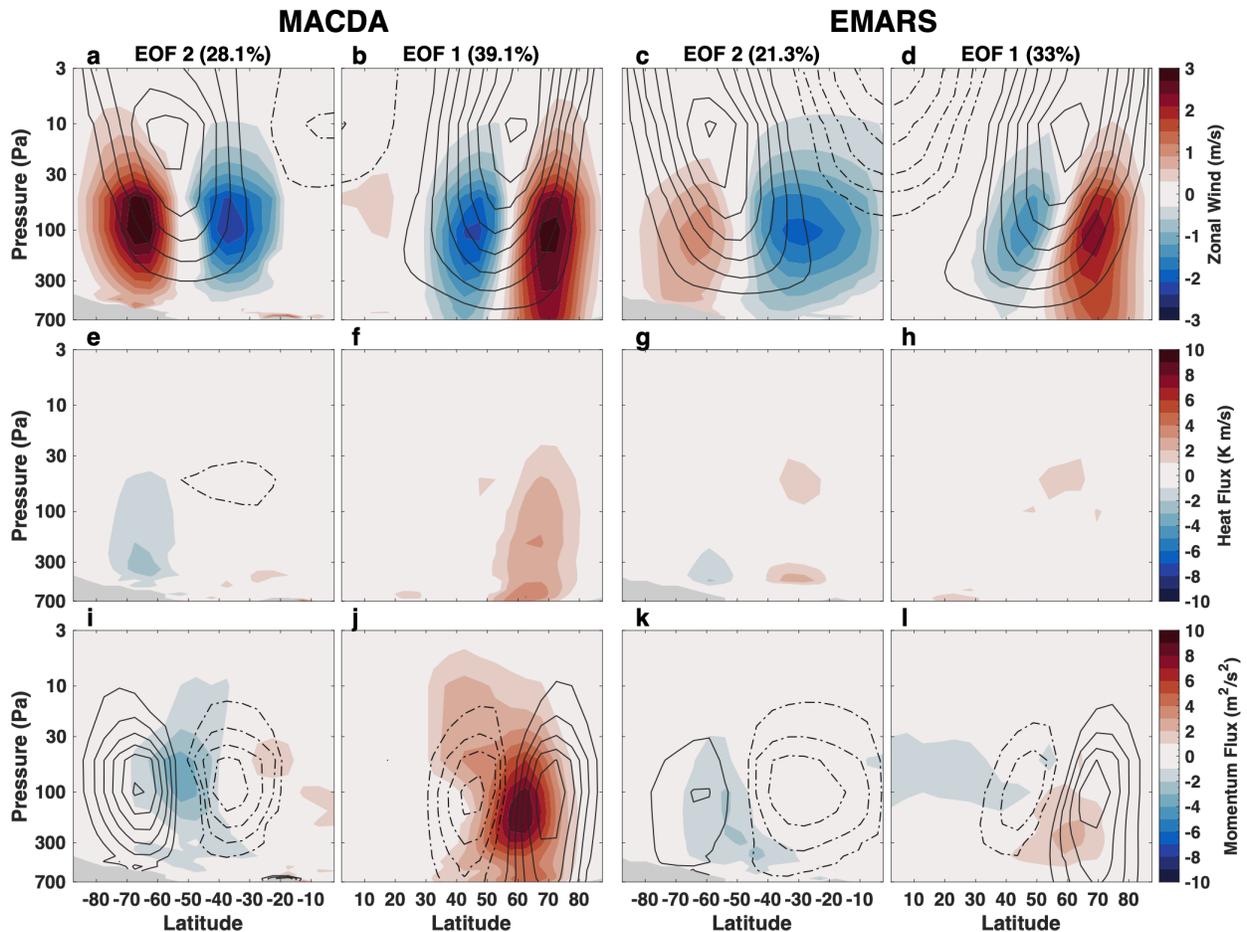

**Figure 2. Spatial signature of the first annular mode in anomalous zonal-mean eddy kinetic energy on Mars for both reanalysis datasets. (a–d)** As in Fig. 1 but for the time-averaged zonal-mean eddy kinetic energy (contours every 200 m²/s²) and regressions onto the anomalous zonal-mean eddy kinetic energy (shading). **(e–h)** Contours duplicated from shading in the top row (contours every 10 m²/s²) and the anomalous eddy heat flux at 0 sol lag (shading). **(i–l)** Regressions of the mode onto zonal-mean zonal wind (contours every 0.5 m/s with dashed contours indicating negative values) and the anomalous eddy momentum flux at 0 sol lag (shading). Regressions are only shown exceeding 99% confidence.

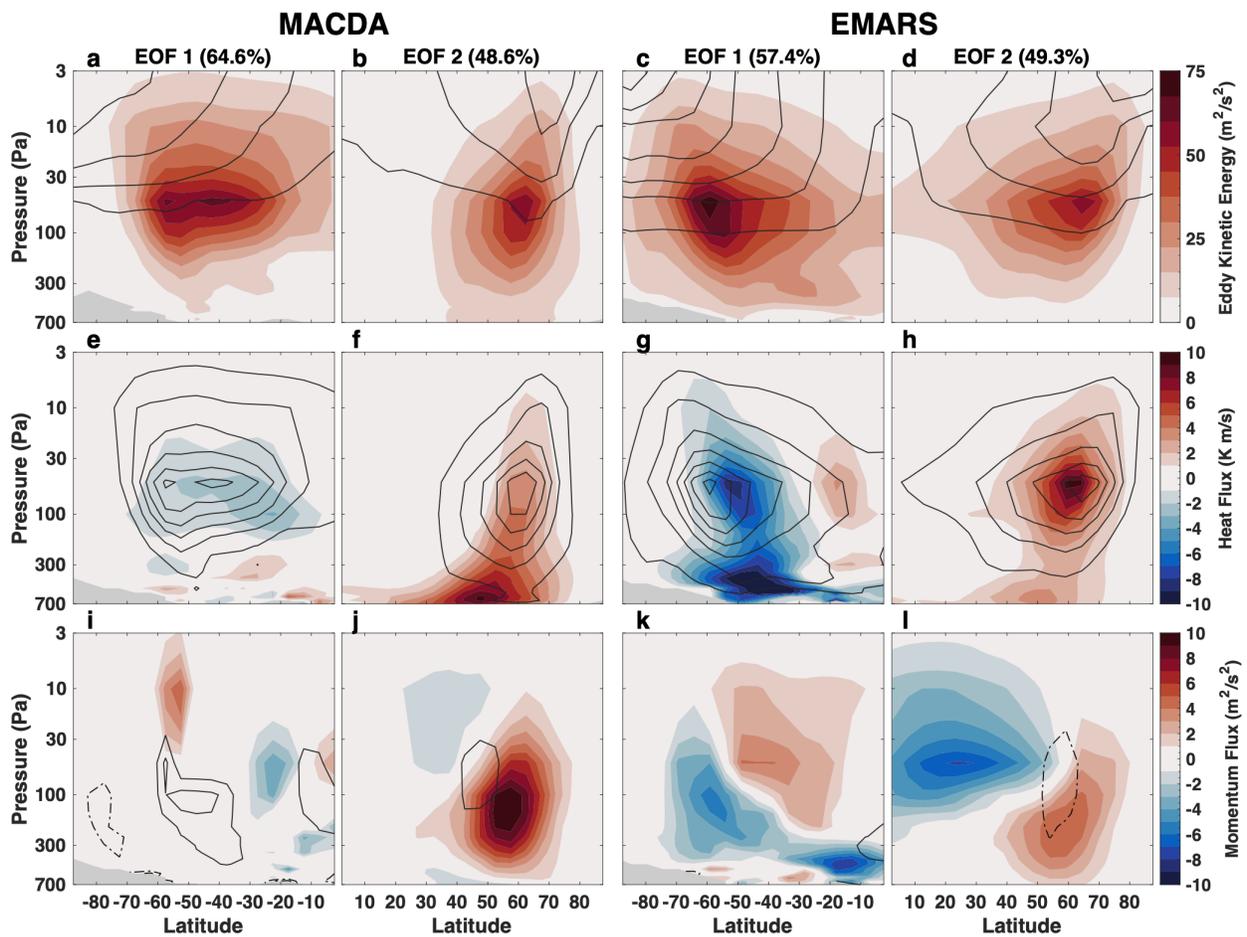

**Figure 3.** Polar plots of vertically (mass) integrated EKE regressed onto the EKE-AM for Earth (a, b), Mars (c, d), and Titan (e, f). (a, c, e) Northern hemisphere. (b, d, f) Southern hemisphere. For Mars, topography is shown in 2000 m increments with the 0 m contour dot-dashed in gray and negative contours dashed. Regressions are only shown exceeding 99% confidence.

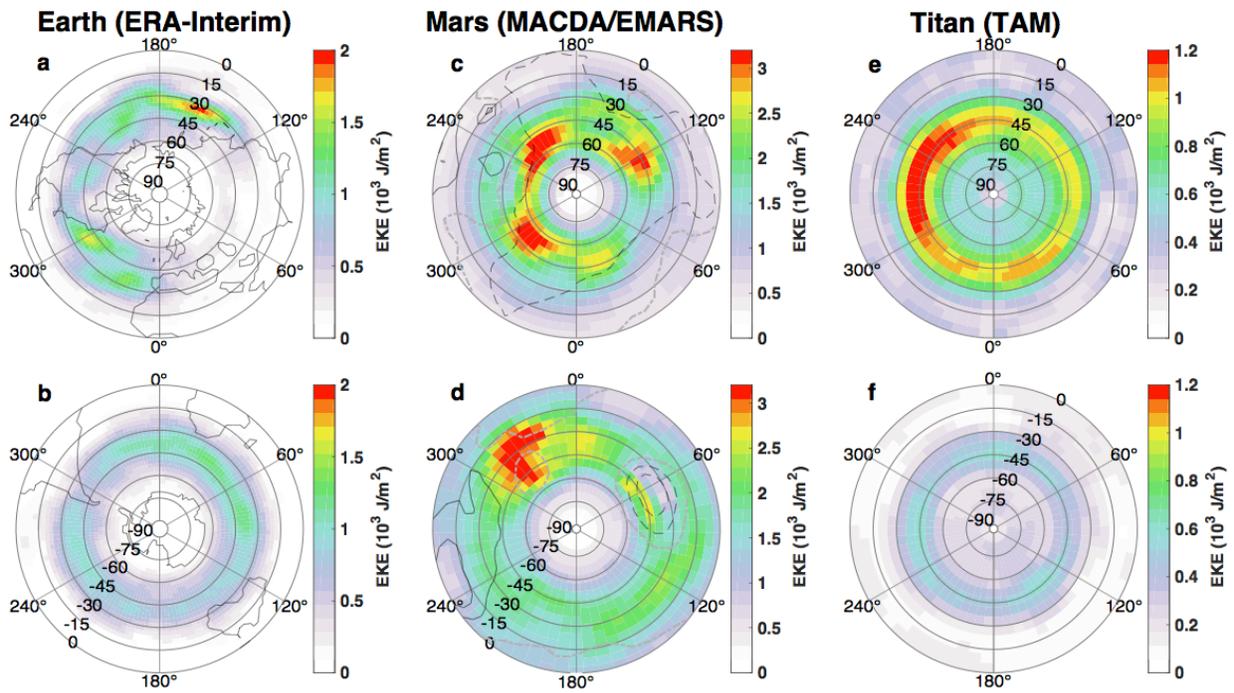

**Figure 4. Regression of the EKE-AM from the EMARS northern hemisphere on the Mars Dust Activity Database for Mars Year 31.** (**a**) dust activity leading the EKE-AM by 4 sols. (**b**) dust activity lagging the EKE-AM by 4 sols. Regressions are performed on the Mars dusty season ($L_s$=180–360º) and are only shown exceeding 95% confidence. Topography is shown in 2000 m increments with the 0 m contour and negative contours dot-dashed.

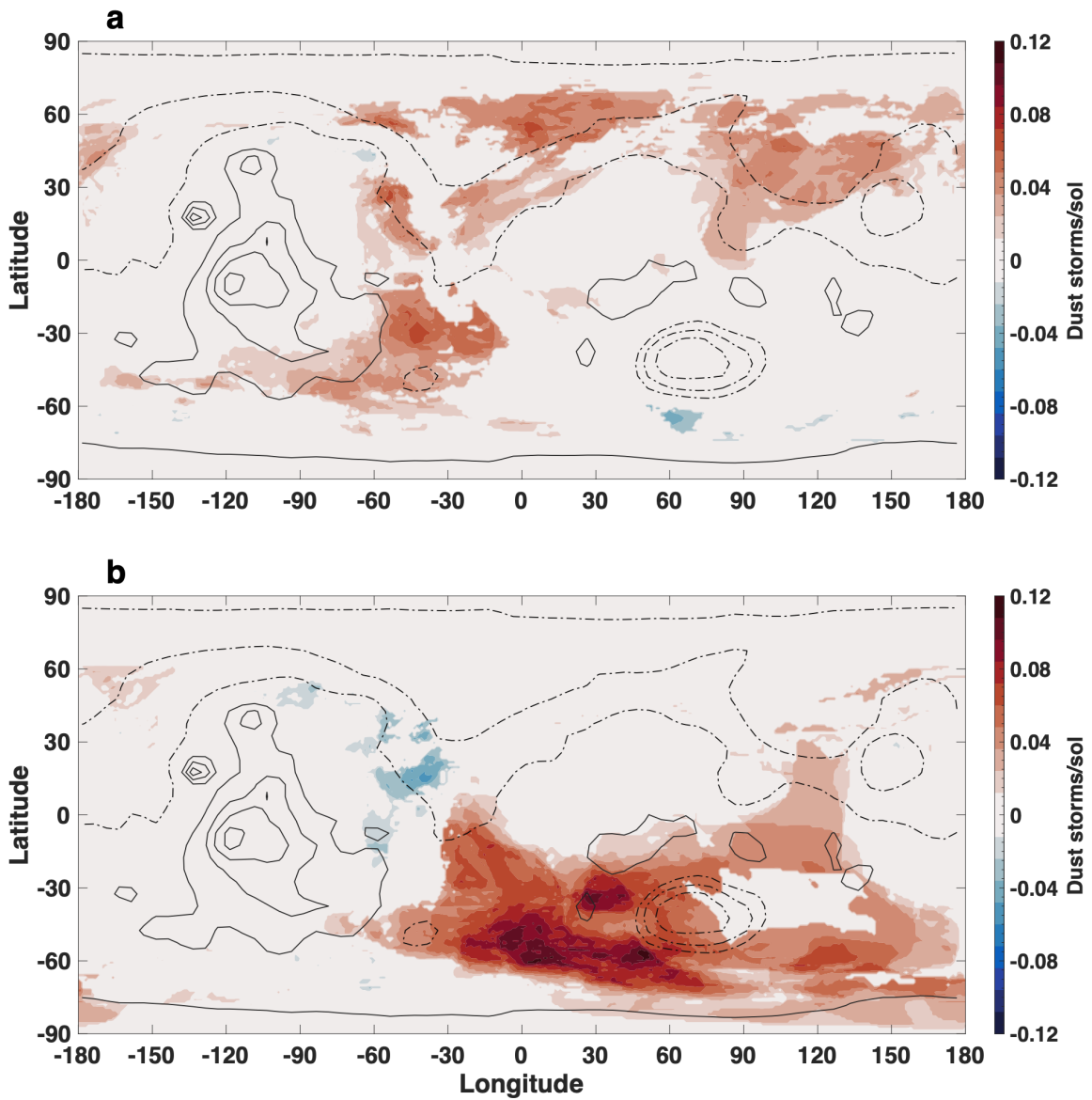

**Figure 5. Zonal-mean structure of the annular modes in zonal-mean zonal wind (left two columns) and eddy kinetic energy (right two columns) on Titan.** **(a, b)** Dataset-averaged zonal-mean zonal wind (contours every 5 m/s) and regression of the U-AM onto the zonal-mean wind (shading). **(c, d)** Dataset-averaged zonal-mean eddy kinetic energy (contours every $5\times10^{-2}$ m$^2$/s$^2$) and regression of the EKE-AM onto the zonal-mean eddy kinetic energy (shading). The individual column titles give the percent of variance explained. **(e–h)** Regressions onto the anomalous zonal-mean eddy kinetic energy (contours every $2.5\times10^{-2}$ m$^2$/s$^2$) and the anomalous eddy heat flux at 0 day lag (shading) for the U-AM **(e, f)** and for the EKE-AM **(g, h)**. **(i–l)** Regressions onto the anomalous zonal-mean zonal wind (contours every $5\times10^{-2}$ m/s) and anomalous eddy momentum flux at -1 day lag for the U-AM **(i, j)** and for the EKE-AM **(k, l)**. Regressions are only shown exceeding 99% confidence.

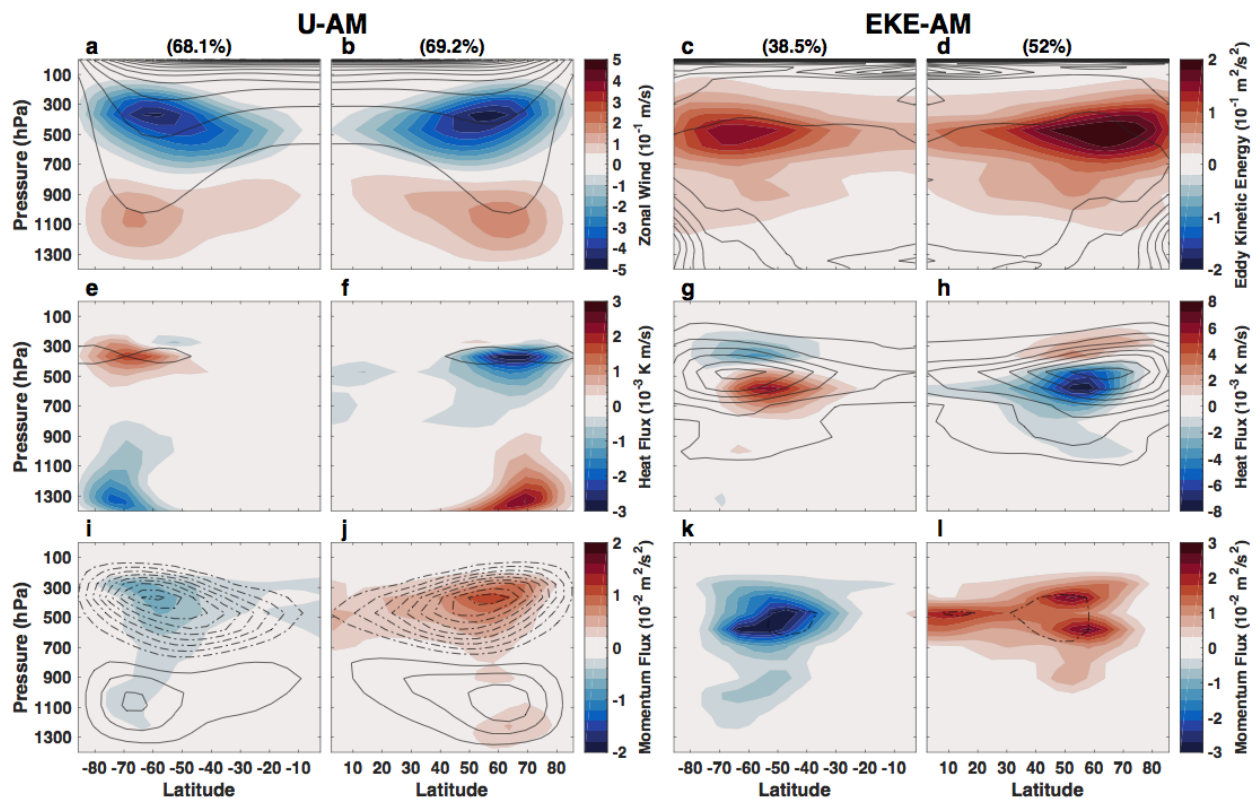

**Methods**

We perform an empirical orthogonal function analysis of annular structures of variability in the atmosphere of Mars using two reanalysis datasets, the Mars Analysis Correction Data Assimilation (MACDA)[29] and the Ensemble Mars Atmospheric Reanalysis System (EMARS). The latter consists of two eras, during which Thermal Emission Spectrometer data and Mars Climate Sounder data are assimilated, respectively[30]. We conduct a similar analysis for Titan using a 20 Titan-year-long simulation of the Titan Atmospheric Model (TAM)[31]. We use the ERA-Interim reanalysis for Earth data[46].

*Time of Year on Mars*

Timekeeping on Mars is related to the position of the planet in its orbit. The seasons on Mars are delineated using the areocentric longitude, $L_s$, which ranges from $L_s = 0$–$360°$. For the northern hemisphere, vernal equinox is $L_s = 0°$, summer solstice is $L_s = 90°$, autumnal equinox is $L_s = 180°$, and winter solstice is $L_s = 270°$.

*Mars Reanalysis Datasets*

MACDA (v1.0)[29] is a reanalysis of Thermal Emission Spectrometer retrievals[47] from the Mars Global Surveyor during the period $L_s = 141°$ MY 24 to $L_s = 86°$ MY 27. Thermal profiles up to 40 km twice per sol and total dust opacities once per sol are assimilated into the UK version of the LMD Mars Global Circulation Model[48] using an analysis correction scheme[49]. MACDA uses a $5° \times 5°$ horizontal grid with 25 sigma levels every two Mars hours.

EMARS (v1.0)[30] includes Thermal Emission Spectrometer data, as well as assimilations from the Mars Climate Sounder[50] onboard the Mars Reconnaissance Orbiter; thus, EMARS spans from $L_s$ = 103º MY 24 to $L_s$ = 102º MY 27 using the Thermal Emission Spectrometer, and $L_s$ = 112º MY 28 to $L_s$ = 105º MY 33 using the Mars Climate Sounder[30]. EMARS is provided at 6º longitude × 5º latitude horizontal resolution with 28 hybrid sigma-pressure levels. EMARS uses a Local Ensemble Transform Kalman Filter to assimilate observations[51,52].

EMARS is an ensemble dataset, meaning that the model is run multiple times with different parameterizations to characterize the "true" synoptic state of the atmosphere as observed. Thus, the ensemble mean is used for the analysis. We favor the use of the ensemble mean as opposed to the use of an individual member because we seek to understand the most likely state of the atmosphere as opposed to a deterministic diagnosis from one model. For the Thermal Emission Spectrometer era, the ensemble has been shown to generally converge to a single solution when analyzing transient waves, but for Mars Climate Sounder data, surface features are less constrained[52]. Nevertheless, repeating our analysis on a single ensemble member from EMARS does not change our results (not shown).

*Titan Atmospheric Model*

Though Mars research has benefitted from semi-continuous and regular observations of its atmosphere over several Martian years, Titan's long year and relative distance have prevented continuous monitoring for any substantial length of time: The recent Cassini mission only sporadically observed Titan during its tour of the Saturn system, which covered half of a Titan year.

Given this relative dearth of observational data and a lack of reanalysis products for Titan, the next best option is an observationally benchmarked general circulation model. Therefore, to explore the possibility of annular modes on Titan, we turn to an analysis of simulations of Titan's atmosphere with TAM[31].

A 20 Titan-year simulation is re-initialized from a previously reported TAM simulation[53]—which includes the atmospheric model coupled to a land model incorporating interactive hydrology—using the preferred version with a surface hydraulic conductivity of k=5 x $10^{-5}$ m/s, which matches best to observations of the hydrologic cycle. TAM has been thoroughly vetted against numerous observations of Titan, and its simulated circulation has been shown to be robust[26,27,31,54], as well as favorably comparable to other models of Titan's climate[27,28,55]. Nevertheless, the Titan analyses herein are of a single model, so should be interpreted with caution.

*Empirical orthogonal function analysis*

Earth's annular modes are diagnosed using empirical orthogonal function (EOF) analysis[3,56,57]. We adopt a similar methodology for Mars and Titan to identify the leading patterns of annular climatic variability. EOF analysis decomposes a time series of data into multiple functions (of the same spatial dimensions as the analyzed dataset) that are determined by statistical relationships within the dataset. EOFs are the eigenvectors of the covariance matrix at each grid point and time step. The eigenvalue associated with each EOF eigenvector corresponds to the variance that is accounted for by the EOF. These functions are orthogonal, meaning they most efficiently represent the variance of the entire dataset. Each EOF of the dataset is paired with a principal component (PC) time series (of the same length as the original dataset). This principal component describes the temporal evolution/amplitude of the EOF at every time step of the dataset[3]. The EOFs and associated PCs are ordered such that the first EOF explains the largest amount of variance of the original field; each subsequent EOF explains the largest amount of remaining variance[3].

The EOFs shown in the analysis are tested for significance, defined as their being well-separated from adjacent modes[56]. This is estimated using the formula $\delta\lambda_\alpha = \lambda_\alpha 4(2/N)^{1/2}$, where $N$ is the number of timesteps and $\lambda_\alpha$ is the eigenvalue for each mode[56]. For all reanalysis domains for Mars and for the TAM simulation, all first modes are well-separated from the second modes, and all second modes are well-separated from the third modes.

The annular modes are defined using the zonal-mean zonal wind [u] and zonal-mean eddy kinetic energy $[EKE] = [u^{*2} + v^{*2}]/2$, where square brackets denote the zonal mean, and asterisks indicate departures from the zonal mean. The eddy momentum flux $[u^*v^*]$ and eddy meridional heat flux $[v^*T^*]$ are also considered. Eddies and fluxes are calculated at each output time step within MACDA or EMARS and then averaged over each sol. Eddies and fluxes are calculated within TAM at each model time step (600 seconds) and averaged over an output frequency of approximately 0.9 Titan days.

For the Martian U-AM, we differentiate spatial structures as dipolar or non-dipolar modes. The non-dipolar U-AM on Mars (Extended Data Fig. 2) explains marginally more variance in the southern hemisphere than the dipolar mode (Fig. 1). The additional mode is tri-polar in MACDA (Extended Data Fig. 2a, b) signifying splitting of the jet and mono-polar in the southern hemisphere in EMARS (Extended Data Fig. 2c) representing intensification of the jet. These variations in structure for both the first and second EOFs are not uncommon for Earth[8,12] and may represent differences in the representation of the jet between reanalyses. The regressed momentum fluxes are colocated with a center of regressed zonal wind (Extended Data Fig. 2i–l). This does not correspond to a barotropic mode and is therefore not considered further.

*Anomalies defined from the seasonal average*

In most studies of Earth's annular modes, EOF analysis is performed on anomalies that are defined by subtracting the seasonal cycle from each year. For this method to successfully approximate the real seasonal cycle, the dataset must be of a long enough duration that a single year with large anomalies does not influence the averaged yearly cycle. For Mars, an analysis performed on anomalies defined from a seasonal average is dominated by the effects of the global dust storms in MY 25 and 28 and to a lesser extent the regional storms that occur each fall and winter[41] (not shown). This induces anomalies that are not real being defined in the years without a global dust storm. The only way to ameliorate the issue is to approximate the seasonal average by filtering out long-period temporal signals using a low-pass filtered time series, to balance capturing the seasonal cycle with removing shorter-period perturbations. A Hamming window filter with a cutoff of 100 Mars or Titan days is used in our analyses. For the EMARS dataset, using the seasonal cycle on the years without a global dust storm (MY 29–33) compared to the full climatology using the filtering procedure provides similar results (not shown) with a correlation of the PC and EOF to the full MY 28–33 EMARS dataset $r$=0.998. Combined with the finding that the spatial patterns of variability during global dust storms are highly correlated to the patterns without global storms ($r$≥0.95), this indicates that global dust storms merely amplify existing spatial patterns of annular variability instead of generating additional modes (see below on *Regression of Principal Components*).

*Sensitivity to domain size*

For Mars, we analyze the daily mean of each variable over the domain 700–1 Pa and 0–90º N/S for the [EKE] and 18–90º N/S for the [u]. The EOF structure of the EKE-AM for Mars is robust as domain size is changed. The spatial structure of the Martian U-AM remains unchanged when the domain size is decreased; however, if the domain of analysis is increased to the equator, an additional mode is revealed within the EMARS datasets that represents strengthening and weakening of the retrograde jet at the equator, which is stronger for large portions of the year in EMARS compared to MACDA[32]. For Titan, we analyze fields from the surface (around 1450 hPa) to the top of the model domain at 0.05 hPa and over the range 8–90º N/S. The TAM results are insensitive to changes in the domain size in the meridional direction, spanning up to 30º degrees and to changes in the vertical direction to as low as 100 hPa. For both Mars and Titan, the southern and northern hemispheres are analyzed separately.

*Weighting in the meridional direction*

Before performing the EOF analysis, the Mars reanalysis data are converted to pressure coordinates in the vertical direction. All data are weighted by mass vertically and weighted in the meridional direction by a factor of $\sqrt{cos\phi}$, where $\phi$ is latitude. Previous efforts to describe annular variability in the atmosphere of Mars using the surface pressure, weighted by $cos\phi$, demoted the annular modes of variability to the third EOF[16,17]. The use of an inappropriate weighting ($cos\phi$) in the meridional direction also relegates Earth's barotropic annular mode in geopotential height to the third EOF, while weighting with $\sqrt{cos\phi}$ places the annular mode as the leading EOF[58]. Applying our EOF analysis of surface pressure from EMARS on incorrectly weighted ($cos\phi$) anomalous, daily mean surface pressure also shows the most prominent annular mode in the third EOF (Extended Data Fig. 1f), whereas the appropriately weighted ($\sqrt{cos\phi}$), anomalous, daily mean surface pressure from EMARS yields a regression map with an annular mode in EOF 1 (Extended Data Fig. 1a). The factor $\sqrt{cos\phi}$ is used because variance, which is what is assessed in EOF analysis, is a squared quantity[58] (see Extended Data Fig. 1a versus 1b).

One complexity for Mars is the deposition and sublimation of the $CO_2$ ice cap, which breaks the relationship between the shift of the zonal wind maximum and mass: for the times of year where the $CO_2$ ice cap is changing, atmospheric mass is not conserved. Indeed, the second spatial structure of the U-AM defined from the surface pressure is not annular (Extended Data Fig. 1b), despite the equivalent structure being annular when defined from the zonal-mean zonal wind (not shown). Thus, care must be taken when comparing the U-AM calculated from either the winds or the surface pressure.

*Regression of Principal Components*

The spatial patterns produced by EOF analysis indicate the locations of action of the annular modes but do not indicate the locations where the modes most impact variables of interest. To ascertain these links, we regress the anomaly fields (of zonal wind, momentum flux, surface pressure, etc.) onto the associated standardized leading PC to generate maps of the regression. The PCs are standardized by dividing each by its standard deviation. This is preferred because standardized PC time series are unitless, so the regressed maps have the same units as the anomaly field itself [1]. The resulting maps correspond to anomalies in the regressed field that are associated with variations in the PC. We assess the significance of regression coefficients with the *t* statistic. The number of degrees of freedom used in the test of significance is computed from the lag-1 autocorrelation[59]. Throughout the work, the level of significance is noted in the text and figure captions, and results are never reported below the 95% confidence level.

For the U-AM on Titan, we regress the eddy momentum fluxes at a lag of -1 day. For the U-AM on Mars, we regress the eddy momentum and eddy heat fluxes at a lag of -1 sol, following terrestrial results[6]. For the EKE-AM, we regress all fluxes for both Mars and Titan at a lag of zero, as the fluxes maximize coincident with the PCs.

For the Titan results, we regress the entire PC time series onto each field of interest. For the Mars reanalysis datasets, we do not regress during the periods of global dust storms (MY 25, $L_s$=170–300° and MY 28, $L_s$=260–325°), due to the large, transient impact on wind and temperature fields. To ensure that the annular modes themselves are not simply artifacts of the global dust storms, we have repeated our analyses excluding the global dust storms entirely. This yields five periods of comparison: before the MY 25 global dust event for MACDA and EMARS, after the MY 25 global dust event for MACDA and EMARS, and after the MY 28 global dust event for EMARS. Each of the EOFs and PCs for both annular modes are correlated to the full run of the analysis at $r \geq 0.95$. This implies that the global dust storms merely amplify the annular modes of variability themselves instead of imposing new patterns of variability.

Dust storms flush from the northern to the southern hemispheres during northern autumn and winter[33,35,36]. So, to prevent inter-hemispheric dynamics from impacting the interpretation of the annular modes, we only present results where we regress the PCs for the northern hemisphere for the period $L_s=180-370°$ and the southern hemisphere for $L_s=10-190°$. These periods correspond to the times of the strongest transient wave activity in each hemisphere[60].

*Mars Dust Activity Database*

Observations of Martian dust storm activity are taken from the Mars Dust Activity Database (MDAD)[35,61]. Each Mars Color Imager, Mars Daily Global Map[36] covers 90º N–90º S. The period $L_s=180-360°$, which is typically considered the dust storm season[35], from MY 31 is used.

The MDAD notes all dust storm activity with well-defined boundaries on Mars with area $>10^5$ km² and indicates each storm individually with an ID number. Dust storms with well-defined boundaries are easily identified from Mars Daily Global Maps, with the edges of the dust storms manually outlined[35]. For comparison to the EKE-AM, the MDAD is re-binned from 0.1º × 0.1º resolution to 1º × 1º resolution, and all of the dust storms on each sol are collected together. The resulting array is regressed against the EKE-AM just as other fields taken directly from the re-analysis datasets.

Correspondence and requests for materials should be addressed to J. Michael Battalio (michael@battalio.com).

**Data Availability**

The Mars Analysis Correction Data Assimilation is available at https://catalogue.ceda.ac.uk/uuid/01c44fb05fbd6e428efbd57969a11177. The Ensemble Mars Atmospheric Reanalysis System is available at ftp://ftp.pasda.psu.edu/pub/commons/meteorology/greybush/emars-1p0/data/. ERA-Interim data are available at https://www.ecmwf.int. The Mars Dust Activity Database is available at https://doi.org/10.7910/DVN/F8R2JX. Titan Atmospheric Model results will be archived on Zenodo or similar.

**Code Availability**

The source code for TAM is currently not publicly available. EOF analysis was done in part with the Climate Data Toolbox for Matlab (https://github.com/chadagreene/CDT). Other scripts used in the study can be obtained from the corresponding author upon request.

**Author Contributions**

J.M.B. conceived of the work. J.M.B. performed the analysis and wrote the manuscript, with contributions from J.M.L. J.M.L. ran simulations of TAM.

**Competing Interest declaration**

The authors declare no competing interests.

**Extended Data Figures**

**Extended Data Figure 1. Polar plots of the regression of the first three EOFs onto the anomalous surface pressure from EMARS in the northern hemisphere.** (a, c, e) results performed using weighting of $\sqrt{cos\phi}$. (b, d, f) results using $cos\phi$. The individual panel titles indicate the percent of variance explained in each EOF. Topography is shown in 2000 m increments with the 0 m contour dot-dashed in gray and negative contours dashed. Regressions are only shown exceeding 99% confidence.

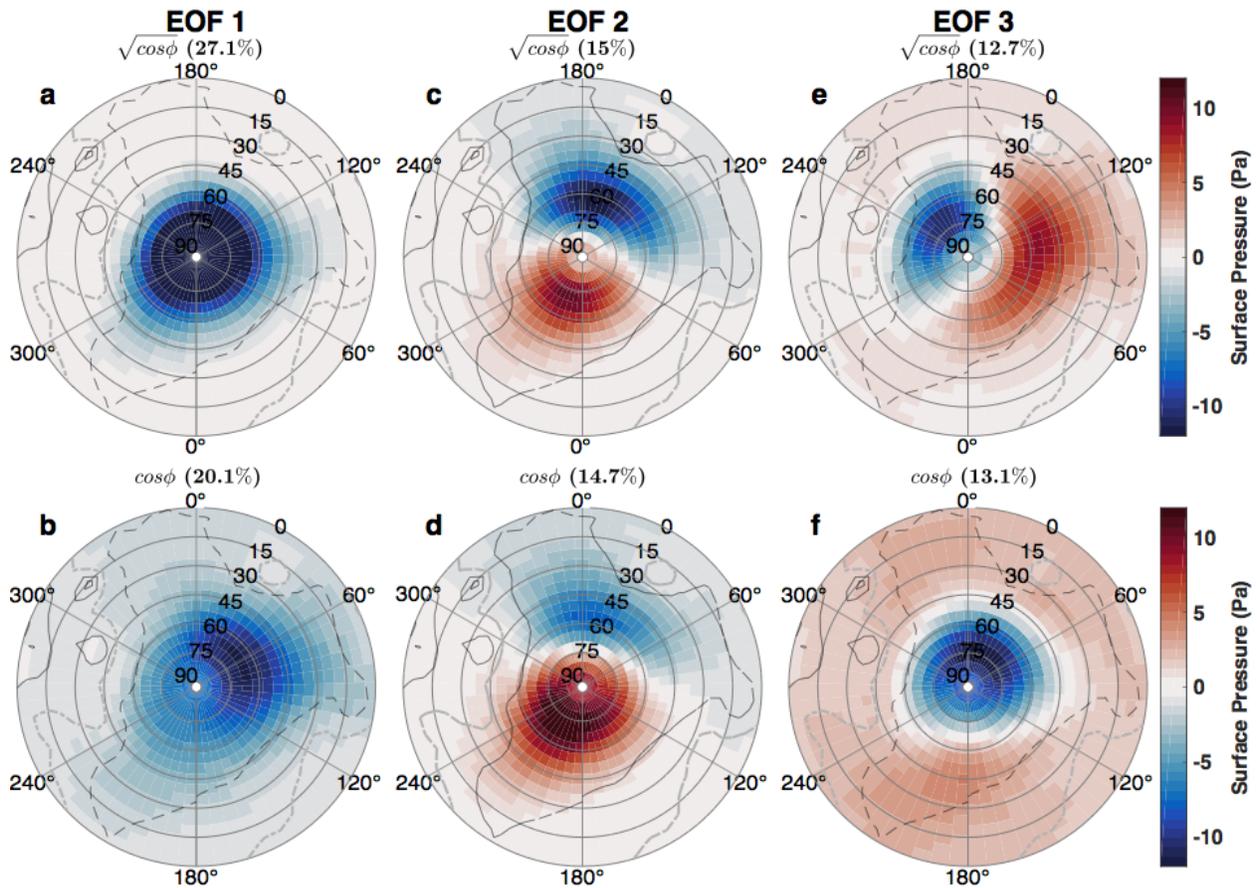

**Extended Data Figure 2. The spatial signature of the first non-dipolar annular mode in anomalous zonal-mean zonal wind on Mars for both reanalysis datasets. (a–l)** As in Fig. 1, but for the first non-dipolar U-AM.

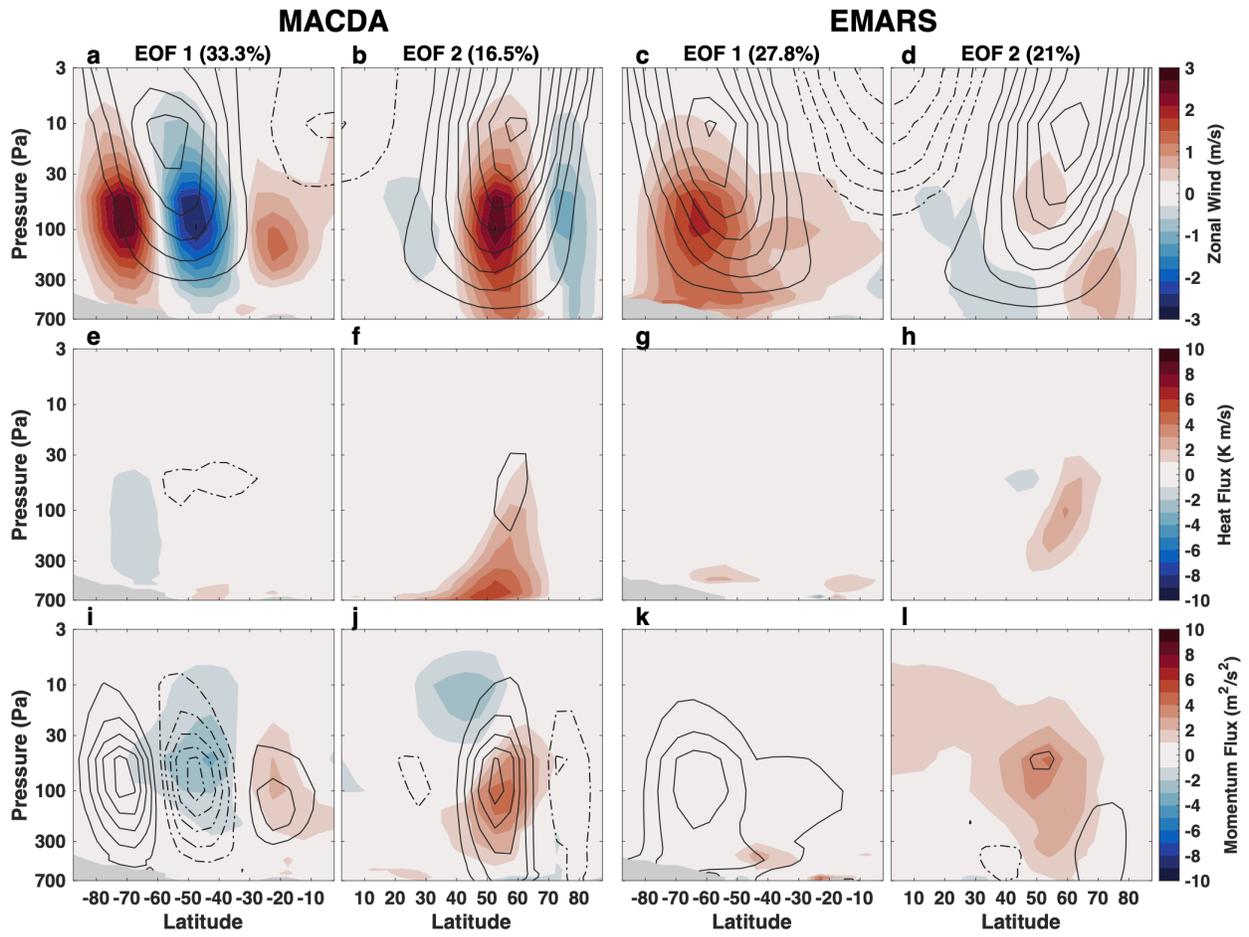

**Extended Data Figure 3.** Polar plots of the regression of the Martian U-AM onto the anomalous surface pressure (a–d) and the regression of the Martian EKE-AM onto the anomalous, vertically (mass) integrated EKE (e–h). **(a, c, e, g)** MACDA. **(b, d, f, h)** EMARS. Topography is shown in 2000 m increments with the 0 m contour dot-dashed in gray and negative contours dashed. Regressions are only shown exceeding 99% confidence.

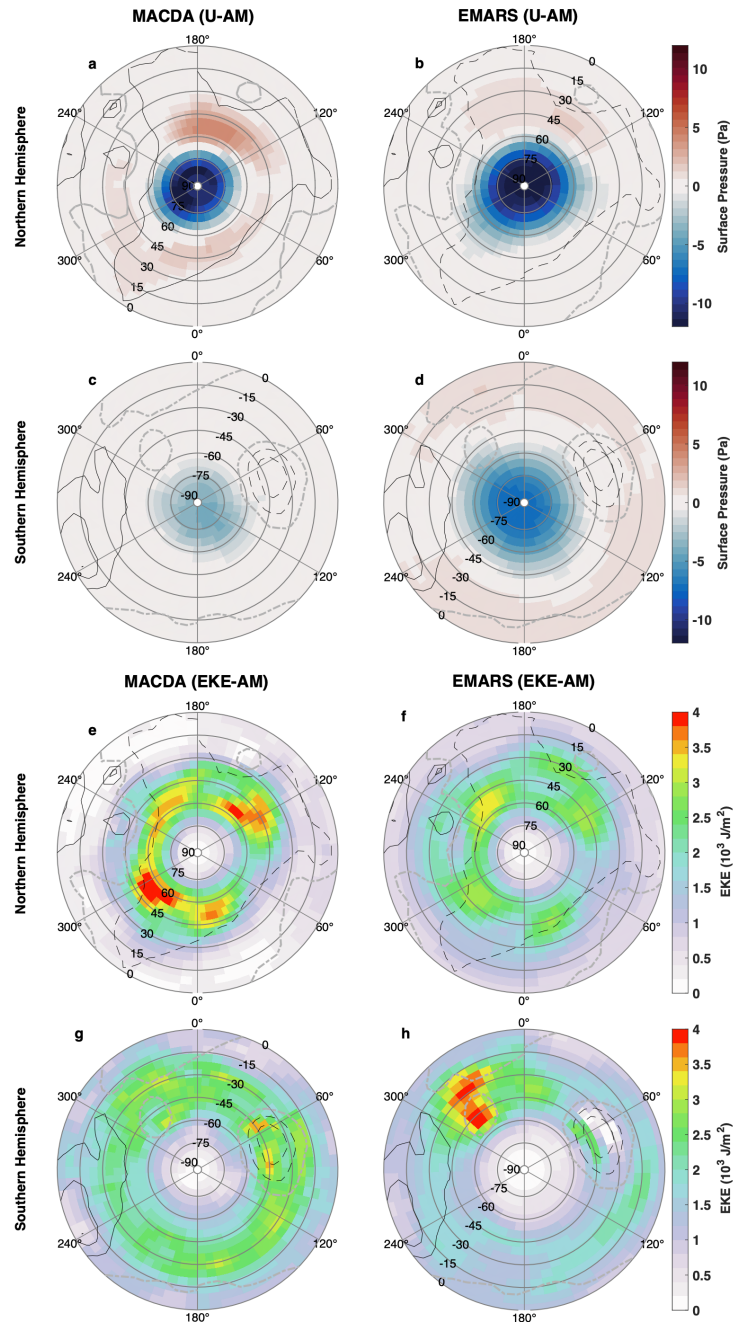